# Magneto Seebeck effect in *RE*FeAsO (*RE*=rare earth) compounds: probing the magnon drag scenario

F. Caglieris [1,2], A. Braggio [2], I Pallecchi [2], A Provino [2], M. Pani,[2,3] G. Lamura [2], A. Jost [4], U. Zeitler [4], E. Galleani D'Agliano [1], P. Manfrinetti [2,3], M. Putti [1,2]

[1] *Department of Physics, University of Genova, Via Dodecaneso 33, 16146 Genova, Italy*
[2] *SPIN-CNR, Corso Perrone 24, 16152 Genova, Italy*
[3] *Department of Chemistry, University of Genova, Via Dodecaneso 31, 16146 Genova, Italy*
[4] *High Field Magnet Laboratory, Radboud University Nijmegen, Toernooiveld 7, NL-6525 ED Nijmegen*

**Abstract**
We investigate Seebeck effect in *RE*FeAsO (*RE*=rare earth) compounds as a function of temperature and magnetic field up to 30T. The Seebeck curves are characterized by a broad negative bump around 50K, which is sample dependent and strongly enhanced by the application of a magnetic field. A model for the temperature and field dependence of the magnon drag contribution to the Seebeck effect by antiferromagnetic (AFM) spin fluctuation is developed. It accounts for the magnitude and scaling properties of such bump feature in our experimental data. This analysis allows to extract precious information on the coupling between electrons and AFM spin fluctuations in these parent compound systems, with implications on the pairing mechanism of the related superconducting compounds.

## 1. Introduction

Six years after the discovery of unconventional superconductivity in high- Tc Fe-based superconductors [1] this fascinating and promising research field is still widely debated, as its origin and fundamental physical mechanisms are yet far from being ultimately clarified. As long as superconductivity appears upon doping of parent compounds, the exploration of electrical and thermo-electrical transport properties of such parent compounds is a powerful tool to address some of the open questions. This task implies disentangling the contributions of several mechanisms, in particular the multiband character and the coupling of the charge carriers to systems of boson excitations such as phonons and antiferromagnetic spin waves. Just these contributions, which may as well play prominent roles in determining pairing interaction and superconducting properties, are responsible for the very complex behavior of transport properties of both the parent compounds and doped superconducting compounds. The most puzzling and articulated among such properties is the Seebeck effect, whose rich phenomenology has been widely investigated from the experimental point of view in iron pnictides of all the families, but still lacks an exhaustive and comprehensive interpretation.

Among the earliest reports of Seebeck effect in iron pnictides of the 1111 family, i.e. with general chemical composition *RE*FeAsO (*RE*=rare earth), McGuire et al. [2] have presented a characterization of Seebeck curves in samples with different *RE*, exhibiting abrupt variations, local maxima and changes in sign. A multiband picture and changes in scattering mechanisms have been proposed to account for the observed behavior. A similar view has been suggested in ref. 3, based on spin density wave fluctuations, which could affect the spin-dependent (possibly also band-dependent) scattering processes, thus causing significant changes in thermoelectric properties. Matusiak et al. [4] have related the large variation of the Seebeck coefficient S in *RE*FeAsO parent compounds below the spin density wave transition $T_N$ to the temperature dependence of the chemical potential. They have also explored the low temperature regime, where *S* curves exhibit a local minimum. The significant sensitivity of such feature to the application of an external magnetic field has suggested to the authors the plausibility of the magnon-drag scenario.

The phenomenology of Seebeck effect in iron pnictides parent compounds of the "122" family, i.e. with chemical composition $AFe_2As_2$ (A=alkaline earth metal), is substantially similar to that of the 1111 family, exhibiting an abrupt change just below $T_N$, changes in sign and a local minimum at low temperature [5,6,7].

In the case of FeTe, considered as parent compounds of the "11" family, Seebeck curves present similar features as the other families such as the abrupt jump below $T_N$ and a local minimum at low temperature, as well as some peculiarities such as the flat temperature behavior above $T_N$ [8,9,10].

In this work, we carry out a careful analysis of the Seebeck effect in the 1111 parent compounds with different *RE*. We explore the dependences on temperature and magnetic field and we propose an interpretative scenario based on magnon-drag by antiferromagnetic spin waves, supported by theoretical models. Within this picture the Seebeck effect comes out to be a privileged property which effectively probes the coupling mechanisms of charge carriers.

**2. Experimental**

The samples were prepared using pure metals and chemical reagents obtained from commercial vendors: the purities were 99.9 wt.% for *RE* (*RE* = rare earth), 99.99 wt.% for As, 99.99 wt.% for $Fe_2O_3$, and 99.5 wt.% for Fe. The synthesis of polycrystalline samples with nominal composition *RE*FeAsO (*RE* = La, Ce, Pr, Sm) was performed by a two-step solid state reaction. In the first step, the *RE*As compound was synthesized and used as a precursor; turning of R and small chips of As were closed under vacuum in a pyrex tube, heated up to, and treated at, 540°C for 3-5 days in a resistance furnace. The second step concerned the synthesis of the quaternary *RE*FeAsO oxy-pnictide. The *RE*As compound, along with the weighed stoichiometric amounts of Fe and $Fe_2O_3$, respectively, were blended and ground together in order to get a homogeneous mixture; the final mixture was then pressed into pellets (total mass of ≈ 2 g, 10 mm in diameter) by using a hydraulic press. The pellets, sealed in outgassed Ta crucibles under an Ar atmosphere, and then closed under vacuum in a $SiO_2$ tube, were subjected to further reaction and sintering in a resistance furnace (1200°C for 4 days); then slowly cooled down to room temperature. *RE*As and *RE*FeAsO compounds were examined by X-ray analysis, using both a Guinier-Stoe camera (Cu K$\alpha$1 radiation, Si as internal standard, $a$ = 5.4308 Å) and a Philips diffractometer (Cu K$\alpha$ radiation). Lattice parameters were calculated from Guinier pattern by means of least square methods, after indexing the patterns.
Seebeck effect measurements were performed from 5 to 300 K and in magnetic field up to 9 T using a Physical Property Measurement System (PPMS, Qantum Design) fitted out with the standard Thermal Transport probe. Seebeck effect measurements up to 30 T were performed at the High Field Magnet Laboratory (HFML) of Nijmegen (NL). All the measurements were performed using a configuration with the magnetic field perpendicular to the gradient of temperature. We performed all the measurements both with positive and negative magnetic fields in order to separate the even part of the signal as respect to the magnetic field, allowing to delete all the odd spurious contributions like Nernst signals. Specific heat measurements were carried out at the Ames Laboratory (US Department of Energy (US-DOE), Iowa State University, Ames, 50011 Iowa, USA) using a PPMS Quantum Design with magnetic field up to 14 T.

**3. Experimental Results**

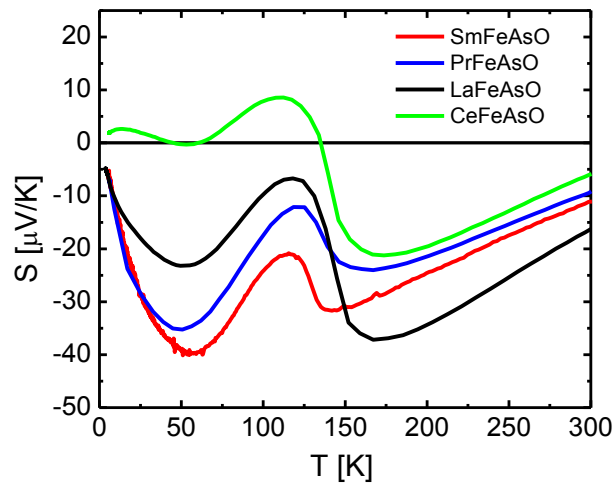

**Figure 1:** Seebeck coefficient curves of *RE*FeAsO (*RE*=Sm, Pr, La, Ce) polycrystals.

In figure 1, we present the temperature dependent Seebeck coefficient curves measured in a series of pnictide parent compounds, having different chemical composition *RE*FeAsO (*RE*=Sm, Pr, La, Ce). It is clearly seen that all the samples exhibit a complex behavior characterized by common features. At high temperature all the curves are negative and decrease in absolute value with increasing temperature. Around 140K, all the curves undergo an abrupt change, related to the magnetic and structural transition. The transition temperature $T_N$ varies between 130K and 145K among these compounds [11]. Below $T_N$ the curves follow different behaviors, before eventually vanishing in the limit of zero temperature. For example, the Seebeck curve of the CeFeAsO sample changes in sign, becoming positive at low temperature, while the other curves are negative. However, even in this low temperature regime, a common feature is observed, namely the presence of a broad bump, responsible for a minimum of S around 50K.

By comparing the reported results with analogous Seebeck effect data measured in 1111 parent compounds [2, 4, 12, 13], it is interesting to note that while the high temperature ($T>T_N$) behavior is largely reproducible, the low temperature behavior is very erratic. In fact, as shown in figure 2 the Seebeck effect at 300K (upper panel) assumes values between -6 and -19 µV/K and it is rather well reproducible for samples with the same *RE*. A very different behavior is observed for the Seebeck effect values at 50 K (lower panel): *S* strongly varies from positive to negative values in the interval between -40 and +20 µV/K, changing from *RE* to *RE* as well as from sample to sample with the same *RE*.

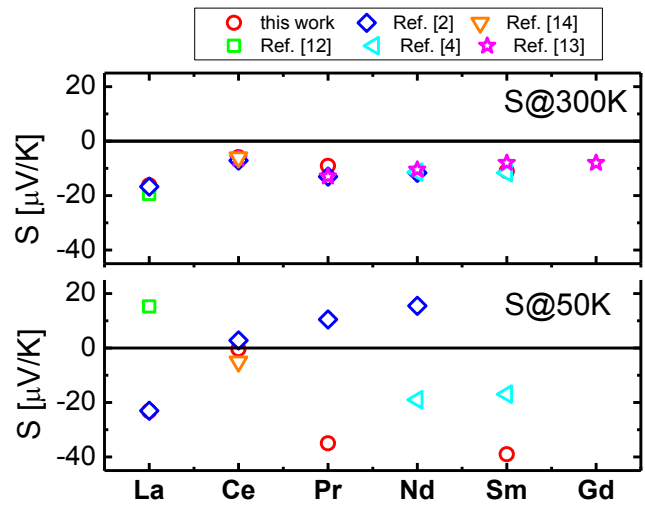

**Figure 2**: Seebeck effect values of *RE*FeAsO compounds at 300 K (upper panel) and at 50 K (lower panel) collected by ref. 2, 12, 4, 13, 14.

In order to emphasize the differences occurring between samples with the same composition, in figure 3 we show the temperature behaviour of the Seebeck effect of LaFeAsO in comparison with data by Kondrat et al. (ref. 12). As it can be seen, above $T_N$ the two curves nearly overlap; on the other hand for $T<T_N$ the curves exhibit opposite behaviors, being our data characterized by a negative bump and Kondrat's data by a rounded positive maximum.

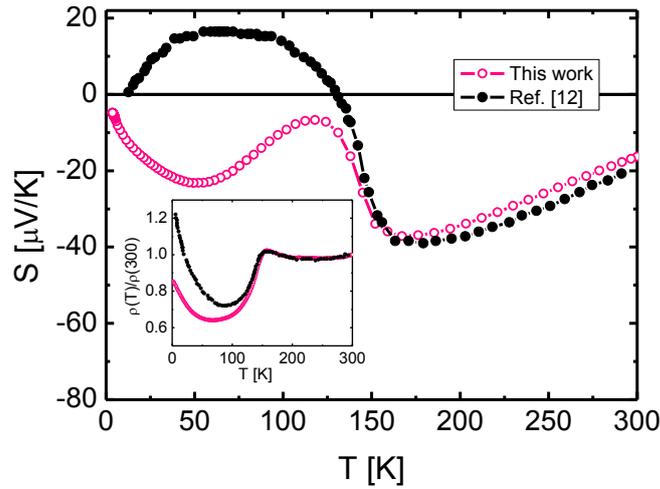

**Figure 3:** Seebeck coefficient curves of LaFeAsO samples: our data in comparison with data taken from ref. 12. Inset: resistivity curves of the same samples, normalized to the room temperature value.

In order to find out the origin of the difference between the two samples, their resistivity curves, normalized to their room temperature resistivity values, are plotted in the inset of figure 3. Above $T_N$, the curves are weakly temperature dependent, and once normalized, they perfectly overlap. They undergo an abrupt dropt at $T_N$ which is followed by a resistivity upturn at low temperature. The latter behavior, indicative of carrier localization, is usually presented by LaFeAsO compounds, at odds with the metallic behavior observed in the case of *RE*FeAsO (with *RE*≠La), and can be related to the lower carrier density of LaFeAsO as compared to other *RE*FeAsO, that emerges from the Hall effect analysis [11]. The resistivity upturn is more evident in the Kondrat's sample than in our own, suggesting that crystallographic disorder responsible for carrier localization could be correlated with the absence of the low temperature negative bump in the Sebeeck effect.

The low temperature behavior of the Seebeck effect shown in figure 3 can be further investigated by exploring the effect of an applied magnetic field. In figure 4, we compare the Seebeck curves of the LaFeAsO sample measured at zero field and at 9T, respectively. Above $T_N$ the two curves overlap, while in correspondence of the low temperature bump they depart significantly, with the in-field curve being larger in magnitude by more than 20%. Similar field dependence has been observed also in SmFeAsO [4] and in LaFeAsO [15]. Interestingly, the Seebeck effect of the Kondrat's sample reported in figure 3 does not depend on the field [16]. Thus, we conclude that the low temperature bump of the Seebeck effect is magnified by the magnetic field and, if the bump is absent, the field dependence disappears.

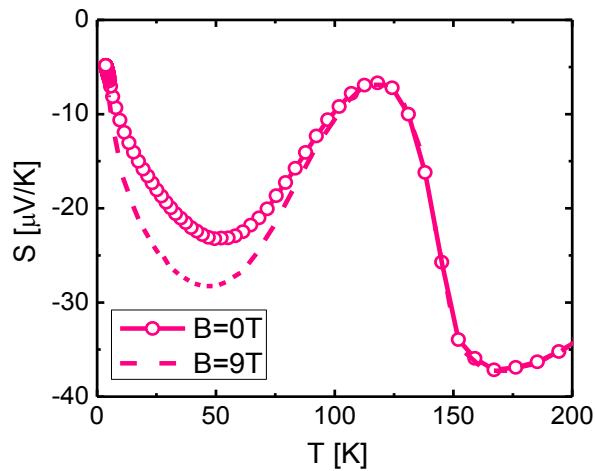

**Figure 4:** Seebeck coefficient curves of the LaFeAsO sample measured at zero and 9T.

A deeper investigation of the field dependence of S is carried out up to 30 T at selected temperatures in the region of the bump. In figure 5 we present isothermal $S(B) - S(0)$ curves versus magnetic field of the LaFeAsO sample performed at T= 30, 45, 60, 77 K. The S absolute values increase in magnitude with increasing field. The overall variation up to 30T is around 50%.

From the results shown so far it turns out that the Seebeck bump is magnetic field dependent, more specifically enhanced by an applied magnetic field and easily suppressed by disorder.

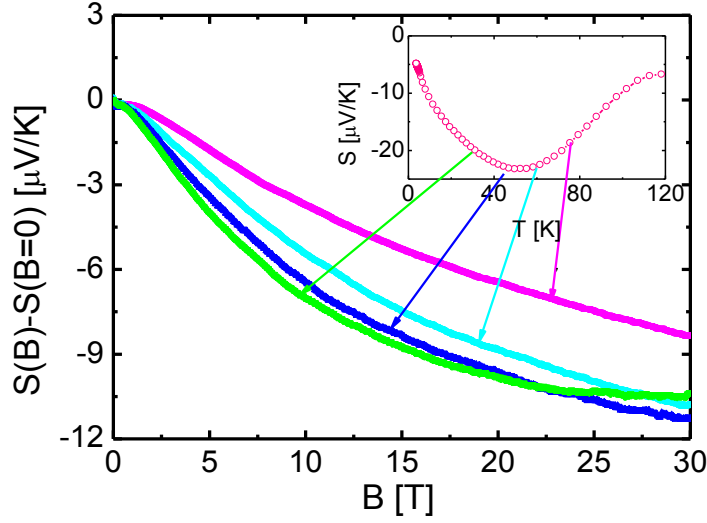

**Figure 5:** S curves versus the magnetic field of the LaFeAsO sample performed at T= 30, 45, 60, 77 K

## 4 Theoretical models

The charge carriers contribute to the Sebeeck effect by different mechanisms. In the following, after recalling the main characteristics of the diffusive and phonon drag contributions, a model for the temperature and field dependence of the magnon drag contribution by antiferromagnetic (AFM) spin fluctuation is proposed.

*Diffusive Contribution*

The diffusive Seebeck effect $S_d$ is due to the motion of charge carriers as a consequence of the thermal gradient. According to Mott formula:

$$S_d = \frac{\pi^2}{3}\left(\frac{k_B}{e}\right) k_B T \left(\frac{d \ln \sigma(E)}{dE}\right)\bigg|_{E_F} \qquad (1)$$

where $k_B$ is the Boltzmann constant, $e$ is the carrier charge with its sign ($e>0$ for holes and $e<0$ for electrons), E is the energy of charge carriers and σ(E) is the spectral conductivity. If we deal with the isotropic case of free electrons scattered by impurities, eq. (1) becomes:

$$S_d = C\frac{\pi^2}{e}\frac{k_B^2}{E_F}T \qquad (2)$$

where the Fermi energy $E_F$, defined positive, is evaluated with respect to the bottom (top) of the band for electron (holes). In eq. (2), C is a dimensionless constant, whose value is 1/3 for a three-dimensional Fermi surface and 1/6 for a two-dimensional one.[17] Considering the expression of the electronic specific heat of a degenerate electron gas with carrier density $n$, $C_e = \frac{n\pi^2}{3}\frac{k_B^2}{E_F}T$, the diffusive Seebeck contribution can be written as:

$$S_d = \frac{C_e}{ne} \qquad (3)$$

Eq. (3) shows that $S_d$ can be interpreted as the average entropy carried by a charge carrier in the material. From this relationship it comes out that in a degenerate single band picture, $S_d$ is expected to follow a linear temperature dependence below $T_F$.

It can be noted that, according to the Mott relationship, the diffusive Seebeck coefficient $S_d$ depends very weakly on disorder. In particular, as σ is proportional to the scattering time τ, $S_d$ includes a logarithmic additive term proportional to $\frac{d\ln\tau(E)}{dE}\big|_{E_F}$ which is almost negligible in most cases, unless $\tau$ is strongly energy dependent. Moreover, as we are well below the magnetic field regime where the electronic structure is substantially affected by Landau quantization, the diffusive contribution to $S$ is not expected to depend appreciably on the magnetic field, either.

*Phonon drag Contribution*

In addition to the diffusive term, the Seebeck effect may exhibit a phonon drag contribution ($S_{ph}$). This term is due to the momentum transfer between the system of phonons and the system of charge carriers and it is observed in the temperature regime where phonons thermalize by scattering preferentially with charge carriers.

A phenomenological expression of the phonon drag contribution is given by [18]:

$$S_{ph} = 3k_B \left(\frac{\alpha_{ph}}{e}\right)\left(\frac{T}{\theta}\right)^3 \int_0^{\theta/T} \frac{x^4 e^{-x}}{(1-e^{-x})^2} dx \qquad (4)$$

where $\theta$ is the Debye temperature and $\alpha_{ph}$ is the effective drag parameter, averaged over the phonon spectrum. This parameter, whose value is in the range 0<$\alpha_{ph}$<1, takes into account the phonon-electron interaction effectiveness and can be expressed as:

$$\alpha_{ph} \approx \frac{\tau_{phe}^{-1}}{\tau_{phx}^{-1} + \tau_{phe}^{-1}} \qquad (5)$$

where $\tau_{phe}^{-1}$ is the phonon scattering rate by electrons and $\tau_{phx}^{-1}$ is the phonon scattering rate by any mechanism other than by electrons (phonon-grain boundary, phonon-defect, phonon-phonon). It can be noted that the limit $\alpha_{ph}\sim1$, that is $\tau_{phx}^{-1} \ll \tau_{phe}^{-1}$, corresponds to the situation where phonon-electron scattering rate is the largest among other relevant scattering mechanisms experienced by phonons. This limit can be fulfilled only in clean samples with large grains and for T<< $\theta$ so that the density of excited phonons is not too large to make phonon thermalization by phonon-phonon scattering dominant. It is easy to verify that $S_{ph}$ can be expressed as:

$$S_{ph} = \frac{1}{3}\alpha_{ph}\frac{C_{ph}}{ne} \qquad (6)$$

Where $C_{ph}$ is the Debye phonon specific heat. For $\alpha_{ph}\sim1$ it turns out that at low temperature, $S_{ph}$ has the same temperature dependence as $C_{ph}$ determined by the temperature excitation of phonon modes, namely ~$T^3$. This behavior is well verified in the normal state of conventional superconductor where the strong electron-phonon coupling makes the condition $\alpha_{ph}\sim1$ more easily fulfilled [19].

At larger temperatures approaching $\theta$, the density of excited phonons increases and the phonons are mainly thermalized by scattering preferentially with other phonons, hence $S_{ph}$ vanishes, exhibiting the characteristic peak around $\theta/5$- $\theta/4$.

At odds with the diffusive contribution, the phonon drag contribution is strongly affected by disorder. Indeed, defects may act as scattering centers for phonons, thus enhancing $(\tau_{phx})^{-1}$ and consequently suppressing $\alpha_{ph}$. In disordered as well as in nanostructured materials the phonon drag contribution to $S$ is hardly observed at all.

On the other hand, similarly to the diffusive term, the phonon drag contribution to $S$ is not expected to depend on the magnetic field.

*Magnon drag Contribution*

Any system of bosons that exchanges momentum with the system of charge carriers introduces in principle a drag contribution to the Seebeck effect in a characteristic temperature range. Hereafter, the drag contribution of the AFM spin density waves is considered. We do not discuss any issue concerning the localized or itinerant nature of these excitations [20] because the present knowledge of the magnon spectrum is still limited for this class of materials yet. Thereby, we assume for simplicity the standard semiclassical approximation for AFM magnons for localized spins [21] in order to extract relevant signatures of magnon drag physics at a general level, taking minimal assumptions on the magnon spectrum. Indeed, as a point of strength, our description addresses the universal signatures of the mechanism rather than the details of the magnon spectrum. Moreover, we keep the number of free parameters at a minimum, also demonstrating that the main results are largely independent from these parameters. We describe the magnons in terms of two branches $E_{\pm}(q)$ ($q$ is the magnon wavevector) corresponding to the spin fluctuations of the AFM ground state. For AFM magnons these two branches may have different gaps, but without experimental evidences of them from literature, for simplicity we assume the same gap $\Delta_0$ for both branches.[22]

Hereafter we focus on the behavior of these branches under an external magnetic field. We have to take account of the vectorial nature of the magnetic field and of the easy axis nature of the AFM order in the considered compound. We need to consider the *longitudinal* and *transverse* contributions, with respect to the easy axis ordering, thus we indicate the projections of the magnetic field along (orthogonal to) the easy axis as longitudinal $B_\parallel$ (transverse $B_\perp$). Indeed the contributions of the two magnetic field components on the magnon branches are different.[23,24,25]

We assume, for simplicity, a completely isotropic gapped magnon spectrum, which, in presence of a magnetic field, can be described by the following analytic expressions:

$$E_+(q,B) = \sqrt{\Delta_0^2 + v^2 q^2} + g\mu_B B_\parallel \quad (7a)$$
$$E_-(q,B) = \sqrt{\Delta_0^2 + (g\mu_B B_\perp)^2 + v^2 q^2} - g\mu_B B_\parallel \quad (7b)$$

where $v$ is the magnon velocity, $g$ the electron Landé-factor and $\mu_B$ the Bohr magneton. For the longitudinal field $B_\parallel$ the two magnon branches are shifted by a Zeeman term $g\mu_B B_\parallel$ in two opposite energy directions. Physically, the external magnetic field helps (contrasts) the creation of magnons in the spin sublattice oriented antiparallel (parallel) to the longitudinal component. In this scheme we also require, for simplicity, that $g\mu_B B_\parallel < \Delta_0$, otherwise the AFM ground state would be modified by the field (spin-flop phase). For the transverse component $B_\perp$ only the $E_-(q)$ branch is modified in the gap term with $\Delta_0^2 \rightarrow (\Delta_0^2 + (g\mu_B B_\perp)^2)$. In conclusion we see that branch $E_+(q,B)$ ($E_-(q,B)$) increases (decreases) in energy with increasing longitudinal field $B_\parallel$, while the presence of a transversal component $B_\perp$ increases only the gap of the $E_-(q,B)$ branch.

We now evaluate the magnon drag contribution to the Seebeck effect. Since we are interested in deriving general properties, we do not solve the problem using a full hydrodynamical approach where a complete analysis of the momentum transfer between electrons, phonons and magnons is taken into account. This kind of approach has been used for phonons with moderate success [26], but it is out of the scope of the present work. Instead, we follow a more intuitive approach inspired by the analysis of the magnon drag Peltier effect carried out for a ferromagnetic (FM) chain [27] and spectacularly confirmed for real cases [28]. The idea is to investigate the contribution to the Peltier effect induced by drifting magnon distributions. Successively, using Onsager symmetry relations we derive the dual thermodynamical quantity i.e. the Seebeck coefficient. This quite direct approach to treat the drag contribution returns a formula that is in fair agreement with the results obtained by more advanced approaches, even if with enormous simplifying assumptions.

We consider a magnon distribution, which is shifted (Galilean translation) by the drag force exerted by the carriers over the magnons through magnon-electron interaction. This corresponds to considering a shift $E_\pm(q,B) \rightarrow E_\pm(q,B) - \hbar v_m q$ with $v_m$ indicating the average magnon drift velocity. This velocity is assumed proportional to the carrier velocity $v_e = j/en$ where $j$ is the carrier current, $n$ the carrier density such that $v_m = \alpha_m (j/en)$. In the latter expression, the drag coefficient $\alpha_m$ is:

$$\alpha_m \approx \frac{\tau_{me}^{-1}}{\tau_{mx}^{-1} + \tau_{me}^{-1}} \quad (8)$$

where $\tau_{me}^{-1}$ is the magnon-electron scattering rate and $\tau_{mx}^{-1}$ the magnon scattering rate with any other relaxing mechanism (magnon-grain boundary, magnon-defect, magnon-phonon, magnon-magnon), such that the denominator in eq. (8) represents the total scattering rate for a magnon. The magnon drag parameter is akin the phonon drag parameter described by eq. (5). The two magnon distributions

$n_\pm^{drift} = n_{\pm 0}[E_\pm(q,B)] + \delta n_\pm(v_m)$ can be written in terms of the stationary bosonic magnon distributions $n_{\pm 0}[E_\pm] = \left(e^{\frac{E_\pm}{k_B T}} - 1\right)^{-1}$ and with the variation $\delta n_\pm(v_m) = -\hbar(v_m q)(\partial n_{\pm 0}/\partial E)$ in the lowest order in the drift velocity. Assuming cubic symmetry of the crystal, the thermal current associated to the drifting distribution, along the $x$ direction, is easily obtained as:

$$j_x^Q = \sum_{j=\pm} \int_{B.Z.} \frac{d^d q}{(2\pi)^d} \delta n_j(q) E_\pm(q) v_x(q) \qquad (9)$$

where the dimensionality of the magnon spectrum is $d$ and the integration is carried out over the magnon Brillouin zone. The last term in the integral represents the magnon velocity along the $x$ direction, $v_x(q) = \hbar^{-1}(\partial E/\partial q_x)$. Using the definition of Peltier coefficient $\Pi = j_x^Q / j_x$ and the Onsager relation $S = \Pi/T$, we find the following quite general result for the Seebeck coefficient of a single crystal:

$$S_m = k_B \left(\frac{\alpha_m}{en}\right) \sum_{j=\pm} \int_{B.Z.} \frac{d^d q}{(2\pi)^d} q_x \left(\frac{E_j}{k_B T}\right) \left(-\frac{\partial n_j}{\partial E}\right) \left(\frac{\partial E}{\partial q_x}\right) \qquad (10)$$

which is the basic formula required to calculate the drag contribution to the Seebeck effect in the case of AFM magnons. In analogy with eq.(4) $\alpha_m$ is the effective drag parameter averaged over the magnon spectrum. Note that the sign of the expression is the same as that of the charge carriers ($e>0$ for holes and $e<0$ for electrons). $S_m$ is inversely proportional to the carrier density $n$ exactly as the diffusive term and the more akin phonon drag term.

We try now to predict the expected magnon drag in particular cases which represent the limiting form of eq. (10) in simpler and relevant regimes. For temperatures high enough to fulfill the condition $k_B T \gg \Delta_0$, we can disregard the gap assuming that the magnon spectrum is linear $E_\pm(q) \approx v|q|$, as typically considered in the literature [29] for cubic symmetry. In this case and for zero magnetic field we easily recover the familiar result:

$$S_m = \frac{1}{3} C_m \left(\frac{\alpha_m}{en}\right) \qquad (11)$$

where $C_m$ is the magnon specific heat. This equation is consistent with eq. (6) obtained for the phonon drag and similar with the one obtained for FM magnons [27]. For AFM magnons in the temperature regime $\Delta_0 \ll k_B T \ll T_N$, the specific heat is proportional to $T^3$, so that, if we could neglect the temperature dependence of the drag parameter $\alpha_m$, also the Seebeck coefficient would inherit the same temperature scaling. Indeed the $T^3$ behavior has been observed in the low temperature Seebeck effect of AFM Chromium [30]. Note that in the opposite limit $k_B T \ll \Delta_0$ the behavior of the AFM magnons is dominated by the gapped spectrum. So the drag contribution to Seebeck is again approximately proportional to the magnon specific heat, which exhibits an activated temperature behavior $\sim T^{1/2} e^{-\Delta_0/k_B T}$.

For the 1111 family the magnon gap $\Delta_0 = 90 K$ has been evaluated from nuclear magnetic resonance data for the LaFeAsO compound [31]; this value is similar to the values found in the 122 family by means of inelastic neutron scattering [32,33,34]. Clearly, in the temperature range T~30-60K we are not in the condition for linear magnon dispersion approximation and the gap $\Delta_0$ cannot be disregarded, thereby the $T^3$ temperature dependence of $S_m$ must not be expected either.

For evaluating the effect of the magnetic field we assume the spectrum given by Eq. (7). Using those expressions we calculate the magnon drag contribution to Seebeck effect $S_m(T,B)$ from eq. (10) as a function of magnetic field in the case of longitudinal and transverse fields, respectively. It is convenient to define the magneto-Seebeck coefficient:

$$MS(T,B) = \frac{S_m(T,B) - S_m(T,0)}{S_m(T,0)} \qquad (12)$$

which measures the relative contribution of the magnetic field dependence in the magnon drag contribution to the Seebeck effect.

In figure 6 we report the calculated $MS(T,B)$ for the longitudinal (top panel) and the transverse field configurations (bottom panel) with the values of magnon gap $\Delta_0 = 90$ K and magnon velocity $v = 3.5 \, 10^4$ m/s. The different color lines corresponds to different temperatures as indicated in the legend. Note that the normalized drag contribution is increased in magnitude as a function of the longitudinal field especially at low temperatures. It is interesting to note that this AFM magnon drag is a growing function of the magnetic field which is the opposite trend as the one predicted [27] and observed in the FM case [28]. This is consistent with the fact that the longitudinal contribution shifts the $E_-(q,B)$ branch al lower energies making it possible to activate more magnons in contributing to the drag. If we instead look at the transverse contribution (bottom panel) we see that the general behavior is a decrease of $MS(T,B)$ with the magnetic field. Note that the vertical axis scale is four order of magnitude smaller than the one for the longitudinal case. This is consistent with the fact that transverse field slightly increases the gap of one branch and therefore the relative contribution described by $MS(T,B)$ must be much smaller. Therefore the signature of the transverse contribution would be in general negligible in the presence of the longitudinal one and in the following analysis we will safely neglect it and consider only the longitudinal one.

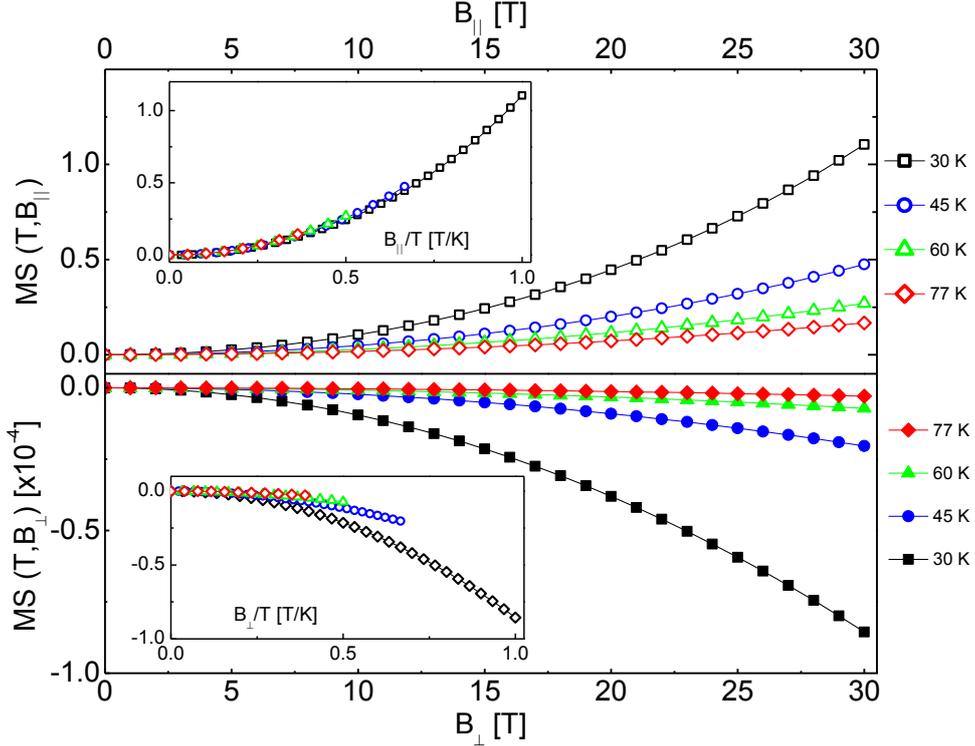

**Figure 6:** $MS(T,B)$ as a function of the magnetic field for the longitudinal field $B_\parallel$ (top panel) and transverse field $B_\perp$ (bottom panel). Note the different vertical scales of the two panels. The different colors corresponds to different temperatures according the top legend. Other parameters are the magnon gap $\Delta_0 = 90$ K and magnon velocity $v = 3.5 \; 10^4$ m/s. Inset: The same data plotted as a function of $B/T$. Note that the longitudinal contribution follows the scaling laws and instead the transverse one does not. See text for details.

It is convenient now to discuss some of the general properties of the drag contribution in the longitudinal case. We consider the limit $k_B T \ll \Delta_0$ [35] assuming the magnon velocity $v \gg a\Delta_0$ with $a$ the crystal lattice constant (which is typically the case for these compounds) and $\alpha_m$ constant in temperature and field. In these limits, eq. (7) (with $B_\perp = 0$) and eq. (10) yield an expression for $S_m(T, B_\parallel)$ that obeys an approximate universal scaling behavior:

$S_m(T, B_\parallel) \sim f(T) g(B_\parallel / T)$  (13)

where the particular functional forms of the functions $f$ and $g$ depend on the details of the magnon spectrum. This peculiar scaling behavior originates from the fact that at low temperatures the activation energy is the parameter which characterizes mostly the magnon spectrum and strongly determines the magnon population and its temperature dependence. The fact that the magnons are subjected to the magnetic field is described by the function $g(B_\parallel / T)$ which is necessarily related to the differential population of the magnon branches. The consequence of this scaling can be nicely observed in $MS(T,B_\parallel)$ which emphasizes the field dependence of the magnon drag cancelling out the important $f(T)$ factor of eq.(13), that contains the temperature dependences of the magnon specific heat and of the drag parameter. $MS(T,B_\parallel)$ as a function of $B_\parallel/T$ is reported in the top inset of figure 6. It is noteworthy that the calculated data in the temperature range (30K-80K) follow quite well the discussed approximate scaling behavior, even outside the strict limit $k_B T \ll \Delta_0$ where the scaling law can be demonstrated to be valid (in the calculation we assume $\Delta_0 = 90$ K). This scaling behavior is far from being trivial. As counterexample we can see that, indeed, it is not obeyed by the transverse contributions as shown in the inset of bottom panel of figure 6.

It is interesting to investigate whether the longitudinal field scaling law is influenced by parameters such as the magnon velocity $v$ and magnon gap $\Delta_0$ and to which extent the scaling law is affected by them. In figure 7a) we report $MS(T,B_\parallel)$ as a function of $B_\parallel/T$, for T=30K and T=60K for of magnon velocity values of $1.4\ 10^4 m/s$ $4.6\ 10^4 m/s$ and for magnon gap values of $\Delta_0 = 90K$ and 70K. It can be seen that the scaling is robust with reasonable values of $v$ and $\Delta_0$. The quantity $MS(T,B_\parallel)$ is virtually insensitive to any reasonable change of these parameters.

Finally, it is useful to consider the dependence of the universal curve of $MS(T,B_\parallel)$ on the g-factor $g$. In figure 7b) we report $MS(T,B_\parallel)$ as a function of $B_\parallel/T$ for $g$=2, 3 and 4 at T=30K and 60K. We can see that with increasing g-factor, $MS(T,B_\parallel)$ correspondently grows [36]. The scaling behavior is still valid but the universal function is affected.

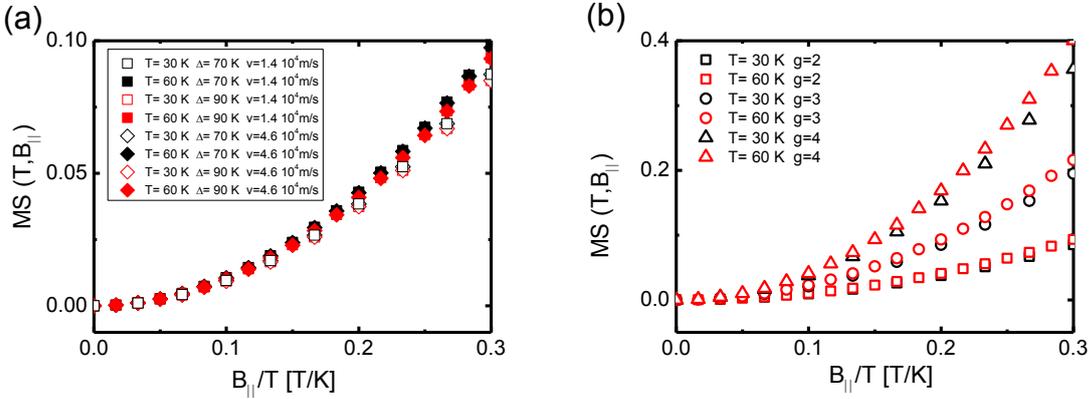

**Figure 7:** (a) $MS(T,B_\parallel)$ as a function of $B_\parallel/T$ evaluated for T=30K and T=60K, magnon velocity values of $v = 1.4\ 10^4 m/s$ (square) and $4.6\ 10^4 m/s$ (triangle) and magnon gap values of $\Delta_0 = 90K$ (filled markers) and 70K (empty markers). (b) $MS(T,B_\parallel)$ as a function of $B_\parallel/T$ evaluated for $g$=2, 3 and 4 at T=30K and 60K, $v = 4.6\ 10^4 m/s$ and $\Delta_0 = 90K$.

We briefly discuss now the limit of validity of our model: eqs. (10)-(13) predict a growing dependence of the magnon drag contribution on the magnetic field, until the critical condition $g\mu_B B_{SF} \approx \Delta_0$ is reached. Indeed above this field a spin-flop transition is expected [37] and the ground-state of the AFM order is modified. We do not expect that this condition is easily reached in our experiment. On the other hand at high field and for high temperatures ($k_B T \sim \Delta_0 - g\mu_B B$) the number of magnons increases enormously and consequently the magnon-magnon scattering rate is expected to increase accordingly. In this condition the drag parameter $\alpha_m$ should be suppressed (see eq. (8)) and the magnon drag coefficient $S_m$ would progressively vanish. This mechanism is the same as the one discussed above for phonon drag coefficient $S_{ph}$ when the Debye temperature $\theta$ is approached. We do not include explicitly this effect in our analysis but it is important to keep into account that it may change the $S_m$ field dependence at high fields. Experimental data may be affected by such mechanism.

Finally we note that the presented analysis is valid for oriented samples, while our experiment has been carried out on polycrystalline sample. In polycrystalline samples each grain has a different orientation with respect to the magnetic field. Even if the transversal component of the magnetic field may be neglected, a directional average of the longitudinal projection should be considered. Furthermore, the refinement of the model should be carried out by taking into account the electronic anisotropy as well, making the analysis more complex and introducing more fitting parameters, which is detrimental to conveying a clear general result. However, we do not expect a change in the discussed scaling behavior even if the polycrystalline nature of the sample is expected to modify substantially the shape of the scaling function of the quantity $MS(T,B)$.

*Multiband effect*

Up to now we have considered the Seebeck coefficients for one band of carriers. As this is not the case of iron-based material, we need to extend the previous results. In the multiband case, the Seebeck effect must be calculated by considering the parallel contribution of all the bands, as the sum of the Seebeck coefficients of each band weighed by the respective electrical conductivities. For two electron and hole bands with conductivities $\sigma_e$ and $\sigma_h$ respectively:

$$S = \frac{\sigma_h |S_h| - \sigma_e |S_e|}{\sigma_h + \sigma_e} \qquad (14)$$

Given that the Seebeck coefficient $S$ of each band is inversely proportional to the carrier density of the band itself, while the conductivity of each band is proportional to the carrier density, it turns out that each term in eq. (14) is independent of the band carrier densities and weighed only by the band mobilities and by other band parameters contained in the expression of $S$ for each type of contribution, such as, for example, the effective masses for the diffusive contribution (see also the following discussion of eqs. (15) and (16)). As a consequence, in a multiband picture the overall temperature dependence of the diffusive $S$ may exhibit very different behaviors, determined by effective masses and temperature dependent carrier mobilities of each band. We will see that to identify which is the most important carrier contribution we need to compare mainly the mobilities rather than the carrier concentrations.

## 5. Data analysis and discussion

As pointed out in the previous sections different contributions to the thermoelectric power should be considered. In particular, the complexity of the curves shown in figure 1 suggests that a competition between different mechanisms must be considered to explain the phenomenology of these compounds.

First of all, we calculate the diffusive contribution in the AFM state. In particular we apply eq. (14) assuming an electron band and a hole band, both having two-dimensional (2D) nature. The 2D nature is motivated by the shape of the Fermi surface of LaFeAsO characterized by quasi-cylindrical electron/hole pockets. In 2D, the Fermi energy expressed in terms of number of carriers is $E_F = \pi \hbar^2 n_{2D}/m^*$ where $m^*$ is the effective mass. Combining this expression with eqs. (2) and (14) we obtain the following compact form for the diffusive Seebeck coefficient:

$$S_d = \frac{\pi}{6c} \frac{k_B^2}{\hbar^2} \frac{(m_h^* \mu_h - m_e^* \mu_e)}{(\sigma_e + \sigma_h)} T \qquad (15)$$

where $c$=8.615 Å is the c-axis of the unitary cell, and $\mu_e$ and $\mu_h$ are the electron and hole mobilities, respectively. The sign of $S_d$ is determined by the factor $(m_h^* \mu_h - m_e^* \mu_e)$. These parameters have been evaluated in the AFM state by magneto-transport properties for the LaFeAsO compound [11]. The values for the hole and electron effective masses taken from *ab initio* calculations [11] are $m_h^*$=0.24$m_0$ and $m_e^*$=0.017$m_0$, indicating a band of very mobile electrons and a band of heavier holes. In fact a ratio of about $\mu_e \approx 10\ \mu_h$ has been evaluated [11], with the mobility values decreasing with increasing temperature. Including these values in eq. (15), $S_d$ turns out to be always positive for T< 100 K.

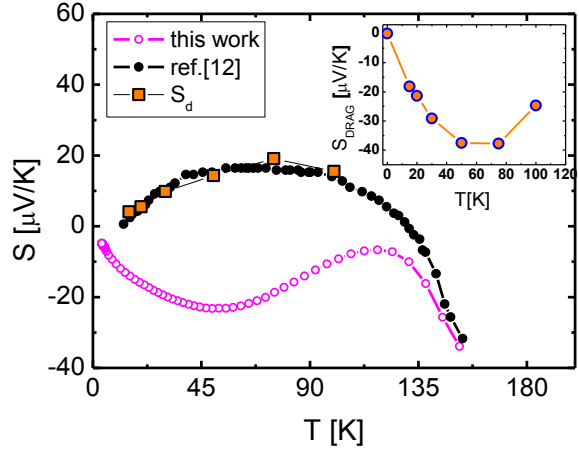

**Figure 8:** Experimental *S* curves of our LaFeAsO sample and of a LaFeAsO sample taken from ref. 12, plotted together with the calculated diffusive contribution $S_d$ (see text). Inset: Drag contribution evaluated by subtracting $S_d$ by the experimental curve as explained in the text.

We point out that the temperature dependence of $S_d$ is not trivial because apart from the explicit linear dependence, also $\mu_e$, $\mu_h$, $\sigma_e$ and $\sigma_h$ depend on the temperature. The obtained values, reported in figure 8, show an initial growing with increasing temperature followed by a broad maximum around 70 K. This behavior appears similar to Kondrat's data, while departs from our data significantly. This comparison strongly suggests that the main contribution present in Kondrat's data is diffusive while our data result from the superposition of the diffusive contribution plus a drag contribution that we identify with the large negative bump around 50 K. Reasonably the drag contribution is washed out in Kondrat's sample by higher crystallographic disorder confirmed by a large value of resistivity at low temperature, already discussed in the section 2.

In order to extract the drag contribution in our sample, we subtract $S_d$ by the experimental data. The resulting curve, $S_{DRAG}(T) = (S(T) - S_d(T))$, plotted in the inset of figure 8, exhibits a negative bump, whose amplitude is maximum at 55K, reaching -37μV/K. This operation provides at least a rough estimation of the drag contribution and its temperature behavior.

In the previous section, two kinds of drag mechanisms are mentioned, caused respectively by phonon and magnon interactions with charge carriers. Distinguishing between these two possible contributions can be difficult in particular if the characteristic temperatures $T_N \sim 150\ K$ and $\theta \sim 200\ K$ [38] are quite close. However the magnetic field dependence shown in figure 4 suggests that magnon drag is the best candidate to account for the Seebeck negative bump.

Thereby, we identify the difference $S_{DRAG}(T)$ evaluated above with the magnon drag contribution $S_m(T, B)$ and we analyse the data on the basis of the scaling argument discussed in the section 5. We consider the following normalized quantity, $MS^{exp}(T, B) = (S(T, B) - S(T, 0))/S_{DRAG}(T)$ where we implicitly assume that only the magnon drag contribution brings the dependence over the magnetic field.

Therefore, we relate $MS^{exp}(T, B)$ with $MS(T, B_\parallel)$ and, as discussed in section 4, we expect that it scales as a function of $B/T$. In figure 8 we plot $MS^{exp}(T, B)$ as a function of $B/T$. As it can be seen, the data at different temperatures virtually collapse into the same curve.

This scaling behaviour is one of the main point of this paper and indeed it is a meaningful finding. Firstly, it is crucial to separate the magnon drag by the diffusive contribution: as shown in the inset of figure 9, where $\frac{\Delta S(T,B)}{S(T,B=0)} = [S(T,B) - S(T,0)]/S(T,0)$ vs $B/T$ is plotted, without this step the scaling would be much less evident. This validates, "*a posteriori*", our subtraction procedure to get the drag contribution. Secondly the scaling essentially validates the magnon drag hypothesis on the basis of quite general assumptions [39].

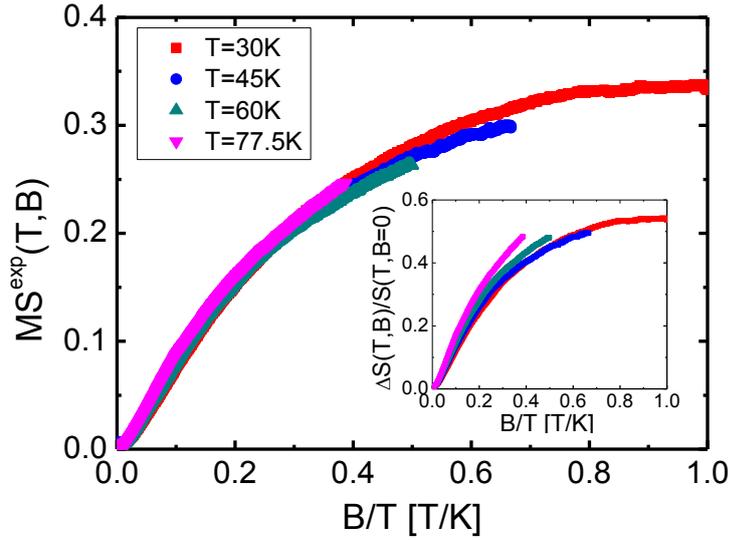

**Figure 9:** $MS^{exp}(T,B) = \big(S(T,B) - S(T,0)\big)/S_{DRAG}(T)$ extracted from the experimental S curves of figure 5 and plotted as a function of B/T. Inset: $\frac{\Delta S(T,B)}{S(T,B=0)} = [S(T,B) - S(T,0)]/S(T,0)$ vs $B/T$.

We point out that $S_{DRAG}$ is negative, which may naively suggest that the electrons, rather than the holes, are strongly coupled with magnons. As a matter of fact, looking at the eq. (11) we can express the contributions of electrons and holes to the magnon drag Seebeck effect as $|S_{m,i}| \sim \frac{1}{3n_{ie}} C_m \alpha_{m,i}$, where $\alpha_{m,i}$ are the effective drag parameters and $i=e, h$ is the band index. Combining the band contributions $|S_{m,i}|$ to the total magnon drag Seebeck effect as in eq. (14) we find:

$$S_m = \frac{\sigma_h|S_{m,k}| - \sigma_h|S_{m,k}|}{\sigma_h + \sigma_e} \sim \frac{C_m/3}{(\sigma_h + \sigma_e)} \big(\mu_h \alpha_{m,h} - \mu_e \alpha_{m,e}\big) \qquad (16)$$

In this expression, the factor $\big(\mu_h \alpha_{m,h} - \mu_e \alpha_{m,e}\big)$ indicates that either high mobility or strong coupling with spin-waves, or both of them, may be responsible for the determination of the sign of $S_m$. In the case of LaFeAsO in the AFM state, the electron mobility is much larger than the hole mobility, $\mu_e \sim 10\mu_h$ [11], which is probably enough in itself in accounting for the negative sign of the drag contribution $S_m$, without invoking stronger coupling with spin-waves of electrons in comparison to the holes.

Coming back to $MS^{exp}(T,B)$ reported in figure 9, it cannot be quantitatively compared with the theoretical model because the sample is a polycrystal and because we assume a constant drag parameter. However, as discussed in the section 4 also in polycrystalline material the magneto-Seebeck effect is expected to grow with $B/T$. Indeed by comparing the theoretical and experimental curves (see figure 6 and figure 9) we see that $MS^{exp}(T,B)$ obeys the expected scaling laws, and also the order of magnitude is roughly comparable with the calculated one.

The discrepancies between theoretical and experimental curves (see figure 6 and figure 9) may have intriguing explanations. The experimental curves show a progressive saturation with increasing $B/T$, while the theoretical curves show only a positive curvature. This may indicate indeed that there is some mechanism that reduces the effectiveness of drag at high field and low temperature.

In our simplified theoretical analysis the field and temperature dependences of the drag coefficient $\alpha_m$ are not considered. Indeed we cannot rule out a dependence on the magnetic field of the magnon-electron scattering rate or, more likely, of the magnon-magnon scattering. We have discussed that with increasing magnetic field the number of magnons increases enormously, potentially making the magnon-magnon scattering dominant and, consequently, reducing $\alpha_m$. It is interesting to note that, being the last mechanisms related to the difference between the magnon populations of the two branches, it is expected to scale with $B/T$, thus any saturation must preserve the scaling.

We wish to conclude by considering the meaning of $S_{DRAG}(T)$. In principle at zero magnetic field we could expect that the Seebeck effect is influenced by all the possible drag mechanisms, namely both phonons and magnons. However, in the previous analysis we assume $S_{DRAG}(T)$ as determined only by the magnons and the scaling analysis supports this assumption. A further confirmation of this hypothesis comes from the comparison of the field dependence of specific heat and Seebeck effect (see Appendix A). Indeed, the observed independence of the specific heat on the field, joined with the strong field dependence of the Seebeck effect, allows to conclude that the drag parameter for the magnons $\alpha_m$ is very large with respect to its phonon counterpart $\alpha_{ph}$, namely $\alpha_{ph} \ll \alpha_m$.

This scenario of strong electron-magnon coupling is remarkable and supports the belief that unconventional superconductivity in *RE*FeAsO systems is mediated by spin waves [40] rather than by phonons [41]. This outcome suggests that Seebeck effect can be viewed as a sensitive probe of carrier interaction providing direct access through the drag contribution to the main coupling mechanism into play.

Our achieved awareness allows to review data in literature on other compounds under a new light. In the 122 parent compounds the Seebeck effect is substantially similar to that of the 1111 family, showing at low temperature a negative bump with features similar to those observed in the 1111 family. The field dependence of the Seebeck effect has not investigated. However, remarkably, Arsenijevic et al. [42] have reported in the $BaFe_2As_2$ a dramatic dependence of the low temperature bump upon application of an external pressure up to 2.5 GPa. As long as pressure has also a significant effect in enhancing the critical temperature of the corresponding superconducting compound, this noteworthy finding offers a clue in establishing a relationship between coupling mechanisms into play, responsible for the magnon-drag enhancement, and active pairing mechanisms, responsible for $T_c$ enhancement in doped superconducting compounds.

The Seebeck effect of FeTe shows an abrupt jump below $T_N$ and a local minimum at low temperature, without the superimposed bump that we attribute to magnon-drag contribution. Noteworthy a virtually negligible field dependence has been measured [8,9]. The missing signatures of magnon-drag suggest that the spin fluctuations related to AFM ordering in FeTe do not couple significantly with charge carriers. This scenario matches with the experimental [43,44,45] and theoretical [46] findings that in FeTe the Fe moments align according to a magnetic wave vector $(\pi, 0)$, in contrast with the AFM order along the nesting wave vector $(\pi, \pi)$ of 1111 and 122 parent compounds. While the $(\pi, \pi)$ spin fluctuations couple with carriers [47,48], $(\pi,0)$ spin fluctuations are not expected to, because they do not match any nesting wave vector [49,50]. This is observed in $Fe_{1+x}Te_{0.7}Se_{0.3}$ superconducting samples where with increasing the interstitial iron concentration (x), $(\pi, \pi)$ spin fluctuations disappear in favor of $(\pi,0)$ spin fluctuations and superconductivity disappears [51]. We predict that in principle also FeTe devoided of interstitial iron should exhibit magnon-drag Seebeck contribution.

We conclude that the magnon drag contribution to the Seebeck effect could return important information over the carrier-spin fluctuation interaction and should be considered for further investigation both in order to further validate the proposed pairing scenario and to extract more quantitative information on the coupling mechanism.

## 6. Conclusions and perspectives

We measured Seebeck effect curves in *RE*FeAsO polycrystals as a function of temperature and magnetic field up to 30T. We observed a remarkable field dependence in the AFM state and we identified different contributions to the Seebeck effect, in particular the diffusive multiband contribution and the magnon drag contribution. The latter was analysed with the support of a theoretical model for the magnon drag in a uniaxial AFM ordered material. We show how the magnon drag contribution depends on the magnetic field and obeys a universal scaling law $\propto B/T$, at least in the regime of our experimental data, once the diffusive contribution is subtracted.

We think that the demonstration of the observed scaling supports the validity of the magnon drag hypothesis but the polycrystalline nature of our samples does not allow to extract reliable information on the specific $(B/T)$ dependence of the drag coefficient.

However, the observed dependence of the Seebeck effect on the magnetic field supports a scenario of strong carrier-spin wave coupling and demonstrates that that Seebeck effect, and specifically its drag contribution, is a very sensitive probe of carrier interaction mechanisms.

The proposed framework must be further tested by investigating samples where the disorder is introduced in controlled amounts, just to have a better check on the diffusive contribution. Finally, the investigation of single crystals could allow to achieve a better knowledge on the spin density wave spectrum trough neutron scattering experiments and thus could open the possibility of extracting more detailed information on the magnon scattering processes and the carrier-spin wave interaction.

**Acknowledgement**

We acknowledge Nicodemo Magnoli, Lara Benfatto and Andrea Amoretti for fruitful discussions.
A. Provino would like to thank Columbus Superconductors SpA, and Regione Liguria for giving her the opportunity to spend abroad a stay at the Ames Laboratory, Ames, Iowa.
We acknowledge the support of the MIUR-FIRB2012 - Project HybridNanoDev (Grant No.RBFR1236VV) , PRIN 2012X3YFZ2 and the EU FP7/2007-2013 under REA grant agreement no 630925 –COHEAT.
This work has been performed at the HFML-RU/FOM member of the European Magnetic Field Laboratory (EMFL) and was partly supported by the EuroMagNET II Project financed by the European Union under Contract 228043.

**Appendix A**

In the previous sections we showed that the Seebeck effect in the AFM region is strongly affected by the magnon-drag contribution. This is clearly a signature of the strong electron-magnon coupling in these materials. In order to gain further insights into this mechanism, we take advantage of the strong correlation between Seebeck effect and specific heat, already discussed in section 3. In particular, for a single band, the following effective relationship can be written:

$$S \approx \frac{1}{nq}\left(C_e + b_{ph}C_{ph}\alpha_{ph} + b_m C_m \alpha_m\right) \quad (A1)$$

where $b_{ph}$ and $b_m$ are numeric constants of the order of unity, whose values depend on non-universal features (such as momentum dependence of the magnon spectrum or energy dependence of scattering mechanism), and $\alpha_{ph}$ and $\alpha_m$ averaged over the spectrum of the excitations. From eq. (A1) it turns out that in the case of drag parameters close to unity, similar temperature and field dependences of S and $C = C_e + C_{ph} + C_m$ are expected. On the contrary the differences between S and $C$ may provide some hints on the drag parameters, and consequently, on the carrier interaction mechanisms.

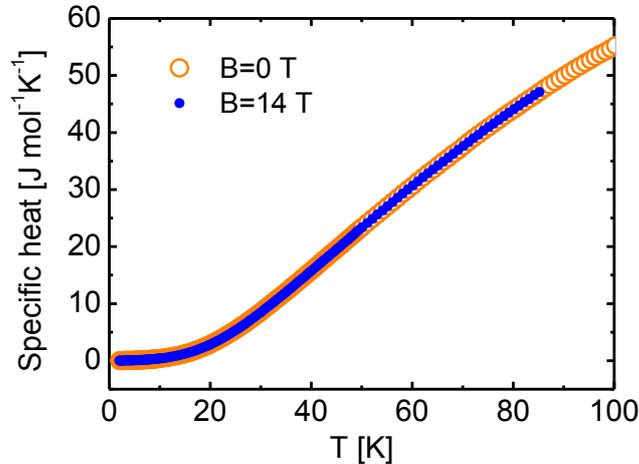

**Figure A1:** Specific heat curves of the LaFeAsO sample measured in zero and 14T magnetic fields.

We thus investigate the specific heat and its field dependence in the temperature range where the magnon drag contribution is observed (10 K <T< 100 K). In figure A1 the specific heat curves of the LaFeAsO sample from 2 to 90 K, both in zero and 14T magnetic field are reported. It is clear that within the experimental sensitivity no field dependence is detected, namely $\frac{C(14T)-C(0)}{C(0)} = \frac{\Delta C}{C} < 0.001$ (experimental sensitivity), whereas any magnon contribution to the specific-heat would be field dependent, as observed in other magnetic systems such as AFM manganites [52] and iridates [53], and quantitatively explained within the spin-wave theory [52]. Thus, assuming that only the magnon contribution to the specific heat $C_m$ depends on the field we estimate:

$$\Delta C_m < 0.001 \left(C_{ph} + C_m\right) \quad (A2)$$

where the electron contribution $C_e$ is neglected for simplicity in the considered temperature range. On the other hand the variation of the Seebeck effect with magnetic field is far from being negligible. Indeed, in the magnon drag regime, at 14 T and 30 K we have approximately (see figure 5) $\frac{S_m(14T)-S_m(0)}{S_m(0)} = \frac{\Delta S_m}{S_m} \sim 0.2$. Thus, using the notations of eq. (A1) we write:

$$\frac{\Delta S_m}{S_m} \sim \frac{b_m \Delta C_m \langle \alpha_m \rangle}{b_m C_m \langle \alpha_m \rangle} = \frac{\Delta C_m}{C_m} \sim 0.2 \qquad (A3)$$

where we assume that the field dependence of $S_m$ is mainly due to the field dependence of $C_m$, as a result of the magnon density. By combining eq. (A2) and (A3), we find an upper limit for the ratio of magnon to phonon specific heats $\frac{C_m}{C_{ph}} < 0.005$. This is the condition the yields simultaneously negligible field dependence of $C$ and large field dependence of $S$. This finding indicates that the phonon density largely exceeds the magnon density, which may appear puzzling if we consider that in section 5(see inset figure 8) it is estimated for T< 100 K $|S_{DRAG}| = S_m \sim 2|S|$, ruling out the presence of any sizeable phonon drag contribution. However, we rationalize both evidences of phonon density largely exceeding the magnon density and of phonon contribution to $S$ much smaller than the magnon drag contribution, by resorting again to eq. (A1), which is compatible with such situation provided that $\alpha_{ph} \ll \alpha_m$.

Furthermore, the observed magnon drag scenario is valid for the minimal value for the ratio $\alpha_m/\alpha_{ph} \gg C_{ph}/C_m \sim 10^2$.

This result supports the conclusion that $RE$FeAsO systems are significantly coupled with spin-waves rather than with phonons, even if, due to the multiband character neglected in the above evaluation, this conclusion is only qualitative.